\documentclass[11pt,twoside]{article}

\usepackage{asp2006}
\usepackage{lscape}

\begin{document}
\title{The Survey of Extragalactic Nuclear Spectral Energies}   
\author{J. A. Gupta and I. W. A. Browne}   
\affil{Jodrell Bank Centre for Astrophysics, School of Physics and Astronomy, The University of Manchester, Manchester M13 9PL, UK}    
\author{B. M. Pazderska}  
\affil{Toru\'{n} Centre for Astronomy, Nicolaus Copernicus University, 87-100 Toru\'{n}/Piwnice, Poland}

\begin{abstract}
We present the first results from the new Survey of Extragalactic Nuclear Spectral Energies (SENSE) sample of ``blazars''. The sample has been chosen with minimal selection effects and is therefore ideal to probe the intrinsic properties of the blazar population. We report evidence for negative cosmological evolution in this radio selected sample and give an outline of future work related to the SENSE sample. 
\end{abstract}

\section{Introduction}   

Traditionally, blazar samples are selected using methods that can introduce severe selection effects. This makes it hard to distinguish between intrinsic sample properties and trends that have been induced by the selection process. Selection effects can therefore lead to erroneous conclusions being drawn about a sample that will hinder progress in understanding the physics of blazars. We have compiled a new redshift-limited sample of ``blazars'', designed to be free from many of the biases that blazar samples tend to incur. 

\section{The Sample}

The SENSE sample contains 160 objects, selected following the method used by \citet{Mar96} for the $200\, \mathrm{mJy}$ sample. Radio and optical data were used to select all sources in the SDSS imaging area that satisfy the selection criteria:
\begin{itemize}
\item compact radio core, $S_{\mathrm{5 GHz}}>90\,\mathrm{mJy}$.
\item redshift $z<0.2$.
\item SDSS optical magnitude $r<18$.
\end{itemize}
The SENSE sample is therefore a volume-limited sample of sources chosen primarily on the strength of their nuclear emission. This selection method avoids many of the selection effects that have plagued past blazar samples. It also means that the SENSE sample contains sources that might not normally be found in blazar samples. In terms of conventional definitions, the sample contains 110 galaxies, 33 BL Lacs, 15 optical AGN and 2 unclassified sources. 

\section{Cosmological Evolution}

The $\langle V/V_{\mathrm{max}}\rangle$ test \citep{Sch68} probes the spatial distribution and hence cosmological evolution of a sample. For the SENSE sample, this test yields the result $\langle V/V_{\mathrm{max}}\rangle = 0.445 \pm 0.026$. This implies negative cosmological evolution at the 2$\sigma$ level; there are more sources nearby than at larger distances, implying that sources meeting the SENSE selection criteria are more numerous now compared to the past.

Previous studies have found different cosmological evolution scenarios for blazars, depending on the selection method and waveband. The negative evolution of our radio selected sample is consistent with that reported for previous X-ray selected BL Lac samples \citep[e.g.][]{Rec01} but at odds with the slightly positive evolution presented in other radio selected samples \citep[e.g.][]{Rec00}. It is possible that the different $\langle V/V_{\mathrm{max}}\rangle$ results for X-ray and radio selected BL Lacs could arise from selection effects, as the majority of samples studied so far are small and contain only the strongest sources \citep{Mar95}. This is an area of ongoing research for the SENSE sample.

\section{Future Work}

Radio observations have been made of several SENSE sources at 610 MHz with the GMRT and at 30 GHz with OCRA-p. The GMRT observations will provide information about the extended emission of these sources which will give a measurement of the intrinsic luminosity. Both sets of observations will also give flux density measurements for the compact cores. These flux densities can be combined with data at other wavelengths (e.g. FIRST, SDSS, Fermi) to plot broadband SEDs. An online database is being developed containing all available data for each source.

The minimal selection effects present in the SENSE sample are well understood, so it can be used to study the statistics of the blazar population. A simple jet model comprising of a fast spine and a slower sheath has been developed. The predictions of this model will be compared to the SENSE sample properties to test if it can explain trends observed in other samples, such as the blazar sequence.

\acknowledgements 

JG acknowledges financial support through an STFC studentship. BP acknowledges financial support from the European Social Fund and Polish Government. The members of the SENSE collaboration are S. Ant\'{o}n, I. W. A. Browne, A. Caccianiga, M. Gawronski, J. A. Gupta, N. Jackson, M. J. M. March\~{a} and B. M. Pazderska.


\begin{thebibliography}{}
\bibitem[March\~{a} et al. (1995)]{Mar95}
March\~{a}, M. J. M., \& Browne, I. W. A. 1995, \mnras, 275, 951
\bibitem[March\~{a} et al.(1996)]{Mar96}
March\~{a}, M. J. M., Browne, I. W. A., Impey, C. D., \& Smith, P. S. 1996, \mnras, 281, 425
\bibitem[Rector \& Stocke (2001)]{Rec01}
Rector, T. A., \& Stocke, J. T. 2001, \aj, 122, 565 
\bibitem[Rector et al. (2000)]{Rec00}
Rector, T. A., Stocke, J. T., Perlman, E. S., Morris, S. L., \& Gioia, I. M. 2000, \aj, 120, 1626
\bibitem[Schmidt (1968)]{Sch68}
Schmidt, M. 1968, \apj, 151, 393
\end{thebibliography}
\end{document}